\documentclass[oldversion]{aa} 
\usepackage{graphicx}
\usepackage{txfonts}
\usepackage{natbib}
\bibpunct{(}{)}{;}{a}{}{,}

\pdfoutput=0

\begin{document}
 
\title{Activity-induced radial velocity jitter in a flaring M
  dwarf\thanks{Based on observations collected at the European
    Southern Observatory, Paranal, Chile, 077.D-0011}}

\author{A.~Reiners
  \inst{1}\fnmsep\thanks{Emmy Noether Fellow}
}


\institute{
  Universit\"at G\"ottingen, Institut f\"ur Astrophysik, Friedrich-Hund-Platz 1, D-37077 G\"ottingen, Germany\\
  \email{Ansgar.Reiners@phys.uni-goettingen.de}
}

\date{Received 26 May 2008 / Accepted 13 March 2009}


\abstract {We investigate the effect of stellar activity and flares on
  short-term radial velocity measurements in the mid-M flare star
  CN~Leo. Radial velocity variations are calculated from 181 UVES
  spectra obtained during three nights. We searched for spectral
  orders that contain very few atmospheric absorption lines and
  calibrated them against the telluric A-band from O$_2$ in the
  Earth's atmosphere. One giant flare occurred during our observations,
  which has a very strong effect on radial velocity.  The apparent
  radial velocity shift due to the flare is several hundred
  m\,s$^{-1}$ and clearly correlated with H$\alpha$ emission. Outside
  the flare, only spectral orders containing the most prominent
  emission lines of H, He, and Ca show a correlation to chromospheric
  activity together with a radial velocity jitter exceeding a few
  10\,m\,s$^{-1}$. We identify a number of spectral orders that are
  free of strong emission lines and show no flaring-related radial
  velocity jitter, although flares occurred as strong as 0.4\,dex in
  normalized H$\alpha$ luminosity. The mean radial velocity jitter due
  to moderate flaring is less than 10\,m\,s$^{-1}$. Strong flares are
  easily recognized directly in the spectra and should be neglected
  for planet searches.}


\maketitle
%

\section{Introduction}

\begin{figure*}
  \center
  \resizebox{.9\hsize}{!}{\includegraphics[]{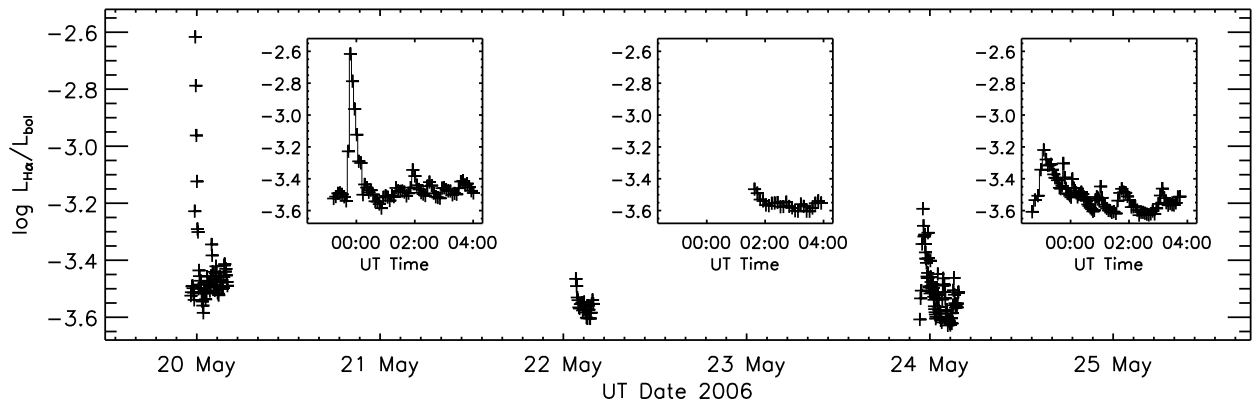}}
  \caption{Normalized H$\alpha$ luminosity during the three nights of
    observation. Time coverage of the three half-nights is shown in
    the plot. The three nights are shown individually in the insets,
    where the variability in H$\alpha$ can be seen. A huge flare is
    clearly visible at the beginning of the first night.}
  \label{fig:Halpha}
\end{figure*}

Radial velocity searches have led to a great many detections of
extrasolar planets. Most of these planets are discovered around stars
of spectral types F, G, and K, although radial velocity amplitudes are
greater in stars of later spectral type and lower mass. The two
reasons for the lack of planet detections around M dwarfs are a) the
intrinsic faintness of M dwarfs render the detection of radial
velocity variations very difficult, and b) many M dwarfs are very
active, which is thought to be a problem for radial velocity
detections, and it is sometimes thought to make M dwarfs less
attractive in the quest for habitable planets.

At the time of writing, 13 planets have been detected around 9 M
dwarfs\footnote{\texttt{http://exoplanet.eu}}. One of them, Gl\,581,
is hosting two planets that potentially fall into the star's habitable
zone \citep{vonBloh07, Selsis07}. Recently, \citet{Tarter07} has
concluded that M dwarfs should be included in the quest for habitable
worlds and evidence of life, because they are so numerous and there
are no obvious reasons why activity should destroy life on them. With
the growing importance of M dwarfs, active or inactive, it is
important to understand the systematic difficulties in detecting
radial velocity variations.

With the radial velocity technique, one detects planets by measuring
radial velocity variations due to the orbital reflex motion of the
host star, which is typically on the order of a few m\,s$^{-1}$
depending on stellar and planetary mass. Radial velocities in general
are determined by measuring the effective blue- or redshift of a
spectrum relative to a reference spectrum. Such a shift in wavelength
can either be imposed by a real velocity variation or by a change in
the appearance of the star's spectrum.  The latter may stem from a
cool spot on the surface of a star effectively covering a part of the
star's surface. Depending on the surface velocity on a rotating star,
a spot can cause virtual radial velocity variations well above the
m\,s$^{-1}$ level. \citet{Desort07} simulated the effect of starspots
on the surface of F-, G-, and K-type stars rotating at moderate
rotation velocities (up to $v\,\sin{i} = 7$\,km\,s$^{-1}$).  They find
that starspots can lead to virtual radial velocity shifts of several
ten m\,s$^{-1}$; a spot covering a fraction of 1.07\,\% of the surface
of a G2 dwarf rotating at $v\,\sin{i} = 3$\,km\,s$^{-1}$ causes a
radial velocity variation with an amplitude of 60\,m\,s$^{-1}$.

The main effect of cool starspots is to remove (or diminish) the flux
coming from the part that is covered with a spot. Cool spots on the
Sun are regions of strong magnetic fields that are several hundred to
a thousand K cooler than their surroundings. The higher atmospheric
layers are heated by magnetic energy so that in the vicinity of cool
spots chromospheric lines like H$\alpha$ become visible in emission.
In very cool stars like M dwarfs, not much is known about the
properties of starspots and their effect on radial velocity
measurements. \citet{Bonfils07} and \citet{Demory07} report radial
velocity variations due to starspots around the planet hosting M
dwarfs Gl~674 and Gl~436. The rotational periods are 35\,d and 48\,d,
respectively, and semi-amplitudes $K_{\rm s} \approx 5$\,m\,s$^{-1}$
were measured. This is roughly consistent with estimates of $K_{\rm
  s}$ from \citet{Saar97}, although their estimates are for a G2 dwarf
with spot temperature zero, which certainly differs significantly from
the situation in M dwarfs.

In very cool stars, the contrast between chromospheric emission lines
and the quiet surface is much higher than in hotter stars so that
observation of activity is easier in cooler stars. A forest of
emission lines becomes visible during strong flares in M dwarfs
\citep[see e.g.][]{Fuhrmeister08}.  The influence of active regions on
the spectra of M dwarfs, however, has not been studied in a systematic
way. One part of the problem is that simulating chromospheric emission
is much more difficult than simulating cool starspots. Nevertheless,
it is expected that strong activity like flares have strong impact on
the radial velocity measurements in M dwarfs.

\citet{Kuerster03} measured radial velocity variations in the inactive
M dwarf Gl\,699 (Barnard's star). They found an anti-correlation
between radial velocity and the strength of the H$\alpha$ line. In the
spectrum of Gl\,699, H$\alpha$ is always seen in absorption. Even in
the ``flare'' spectrum that K\"urster et al. show, H$\alpha$ is not
seen in emission but the absorption is weaker than in the other
spectra \citep[an effect that may actually be due to \emph{lower}
temperature not \emph{higher}; see, e.g.,][]{Short98}.  K\"urster et
al. found a significant anti-correlation between their H$\alpha$
activity level and radial velocity. The small activity variations of
Gl\,699 led to radial velocity variations on the order of
10\,m\,s$^{-1}$. However, the radial velocity measured in the
``flare'' spectrum of K\"urster et al. does not show as high a radial
velocity shift as could be expected from the correlation seen at lower
H$\alpha$ level.

The fact that cool spots on sun-like stars can cause large radial
velocity variations, and the result that the H$\alpha$ line strength
(anti-)correlates with radial velocities even in a very inactive M
dwarf raise the question to what extent activity and flaring events
influence the detectability of real radial velocity variations due to
orbital motion.  This is particularly important since in mid- and
late-M dwarfs the fraction of active dwarfs rises dramatically, and
the majority of mid- and late-M dwarfs show H$\alpha$ in emission
\citep{West04}.

\section{Data}

In a multi-wavelength study of the very active nearby M6 dwarf CN~Leo
(Wolf~359, Gl~406), we obtained high resolution spectra with UVES at
the Very Large Telescope, Paranal, Chile. The purpose of this data was
to investigate the multi-wavelength behaviour of flares. Here, we
investigate this data for radial velocity variations. The target
always shows strong H$\alpha$ emission, and during the three nights it
presented many small flares and one very big one.

We obtained 181 spectra during three half-nights in 2006 (May 19/20,
21/22, and 23/24). The data are described in detail in
\cite{Fuhrmeister08} where a spectral atlas of a spectrum taken during
a huge flare is presented. The original purpose of the data was to
investigate flares, not to search for radial velocity varations.
Hence, no special calibrations were done and the spectra were taken
without the iodine absorption cell. We used a slit width of 1'' ($R
\sim 40\,000$) and the red arm was centered at 830\,nm providing
wavelength coverage from 640\,nm to 1008\,nm in 32 orders. 

Three half-nights were granted to the project. Because of bad weather
conditions, no observations were taken during the first half of the
second half-night. During the other two nights, CN~Leo was observed
more than four hours without interruption. Exposure times varied
between 100\,s and 300\,s depending on weather conditions. Several
flares occurred during our observations, among them a very strong one
that boosted the observed H$\alpha$ emission by more than an order of
magnitude (it saturated the detector so we cannot measure the total
H$\alpha$ emission in that spectrum). The flare is easily visible in
the whole optical region, it is discussed in detail in
\citet{Fuhrmeister08}.

In Fig.\,\ref{fig:Halpha}, normalized H$\alpha$ activity is shown
during the observations. Measured H$\alpha$ equivalent widths were
converted to $L_{\rm{H}\alpha}/L_{\rm{bol}}$ using theoretical PHOENIX
spectra as a reference \citep[see][]{RB07}. Fig.\,\ref{fig:Halpha}
shows the temporal coverage of the data. In the insets, the three
half-nights are shown individually. The strong variability of CN~Leo
is obvious from this plot, and the huge flare during the beginning of
the first night stands out as an extraordinary event.

\section{Radial velocities}

The most successful method detecting extra-solar planets so far is the
radial velocity method. The hunt for radial velocity perturbations
induced by planetary companions became possible when this technique
became sophisticated enough to push the stability of radial velocity
measurements down to several ten m\,s$^{-1}$, i.e. typically only
1/100 of a pixel and less. To reach such an accuracy, wavelength
calibration data is included in the target spectrum itself, so that
any instrumental effects are directly visible during the observation.
One way to reach this is to obtain the spectrum of a calibration lamp
next to the target spectrum \citep[the ``ThAr method'',
see][]{Pepe02}. A second method uses an absorption cell in front of
the spectrograph.  Light from the star passes the cell, which imprints
a dense forest of absortion lines onto the spectrum \citep{Butler96}.
The crucial advantage of both methods is that every systematic effect
in the wavelength solution of the spectrum is directly monitored
during observation (although the ThAr method requires much higher
stability).

Our spectra were not taken with special care regarding to wavelength
stability. For calibration of UVES spectra, a ThAr spectrum is usually
taken at the end of every night. These are the wavelength references
that we use to calibrate our data. The typical stability of UVES,
however, is on the order of a pixel over the course of a night, i.e.
deviations on the order of a km\,s$^{-1}$ can be expected.  In order
to reach an accuracy better than the absolute stability of the UVES
spectrograph, we employ the absorption lines from Earth's atmosphere.
They provide a reliable reference frame on the order of the wind
speed, which does not vary abruptly during a night
\citep{Balthasar82}. None of the telluric lines were removed from the
spectra prior to analysis.

\subsection{Relative radial velocity shifts}

\begin{figure}
  \resizebox{.97\hsize}{!}{\includegraphics[]{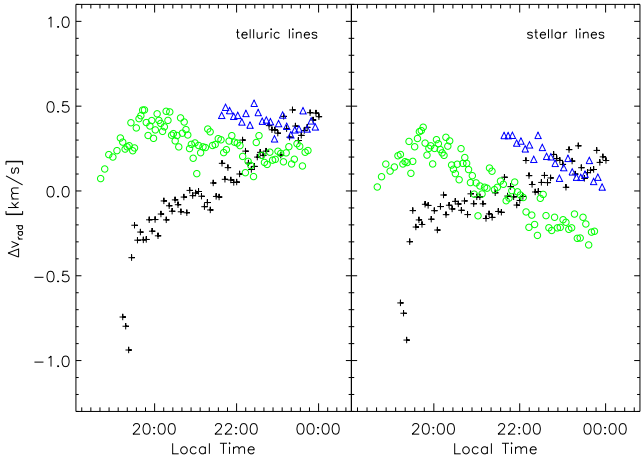}}
  \caption{ \label{fig:vrad_raw}Relative radial velocities $\Delta
    v_{\rm rad}$ between the 181 spectra and an arbitrary reference
    frame (\#24) versus local time at Paranal. Black crosses are data
    from the first night, blue triangles are from the second night,
    and green circles are spectra taken during night 3. Left: Radial
    velocities calculated in the telluric A-band; right: $\Delta
    v_{\rm rad}$ calculated from order \#69 that contains mainly the
    TiO IR band (see text).}
\end{figure}

To measure radial velocity shifts from our spectra, we first
calculated the cross correlation functions relative to a reference
spectrum. Spectrum number 24 obtained during the first night was used
as reference spectrum; it exhibits relatively little H$\alpha$
emission. This was done for all spectral orders individually, i.e. we
calculated 181$\times$31 individual radial velocity differences.  In
the following, these are labelled the \emph{relative} radial velocity
shifts (relative to an arbitrary zero point). They are not corrected
for instrumental effects that may have occurred during the course of
each night.

Fig.\,\ref{fig:vrad_raw} shows all 181 relative radial velocities
$\Delta v_{\rm rad}$. Black crosses are data taken during the first
night, blue triangles are taken during the second night, and green
circles are observations from the third night. In the left panel,
$\Delta v_{\rm rad}$ from order \#80 (757--769\,nm) is shown.  This
order covers the A-band due to atmospheric $\rm O_2$. Assuming no or
very little variations in the telluric lines, $\Delta v_{\rm rad}$
from this order shows the absolute drift of the spectrograph during
the nights. In the right panel of Fig.\,\ref{fig:vrad_raw}, we show
radial velocity differences calculated from spectral order \#69 that
contains the IR TiO band and only very little telluric absorption. No
barycentric correction was applied to the data shown in
Fig.\,\ref{fig:vrad_raw}.

Clearly, relative radial velocities of telluric lines and stellar
lines follow the same gross pattern. During the first night (black
crosses), the telluric and the stellar spectra show a strong redward
shift within the first 15 minutes. More precisely, a blueshift occurs
during the first three exposures followed by a steep redward jump.
After that, a linear redward trend drifting approximately
1\,km\,s$^{-1}$ in four hours occurred. This trend, observed both in
the atmospheric and the stellar lines, clearly is introduced by the
UVES instrument itself and cannot be due to the motion of the star.
The origin of the jump and the linear redward trend during the first
night is unclear, but a thorough analysis of instrumental effects goes
beyond the scope of this paper.  During the second night (blue
triangles) and the third night (green circles), a shallow blueward
trend is observed. At the beginning of the third night, a redward
shift occurred that is similar to the one observed in the first night.
The blueward drift during the second and third nights is on the order
expected from changes due to variable atmospheric pressure inside the
instrument (see next section).

\subsection{Differential radial velocity shifts}

Calibrating radial velocity shifts with telluric lines has several
caveats that must be taken into account. The basic problem is that
telluric lines are spread over almost the entire spectrum, but in
particular in the rich spectra of M dwarfs they are not always easily
discovered. The cross-correlation function of two spectra taken from
the same star generally consists of two components: 1.  The
correlation peak from stellar features, which follow the motion of the
star and the drift due to barycentric motion. 2. The correlation peak
of the telluric lines that are not subject to barycentric velocity
drifts.  If a spectral range contains both sets of features, the
influence of the correlation peak due to telluric lines will affect
the peak from stellar features. In the easiest case, telluric features
dominate and the correlation peak no longer follows the star's motion
but rather the effects of telluric lines.  In a more complicated
situation, the varying distance between the stellar and the telluric
peaks interfere so that the velocity measured from the cross
correlation function is a complex combination of stellar and
barycentric motion.

One way to get rid of telluric features is to correct for them by
using either a telluric standard spectrum or a synthetic set of
telluric lines. Alternatively, one can mask out regions of telluric
lines and use only regions that are virtually free of telluric
contamination. As mentioned above, it is not always straightforward to
identify telluric features in particular in the spectra of M dwarfs.
\citet{Reiners07} show a high resolution spectrum of a standard star
in which telluric absorption at Paranal is visible in the wavelength
region used for this work. This spectrum shows that telluric
contamination is relatively weak outside the strong bands of O$_2$ (A-
and B-band at 760\,nm and 690\,nm, respectively) and the H$_2$O
features at 720\,nm, 820\,nm, and between 890\,nm and 980\,nm. This
means that probably a large range of the spectral coverage of UVES can
be used to measure the radial velocity shift of CN~Leo.

\begin{figure*}
  \resizebox{.97\hsize}{!}{\includegraphics[]{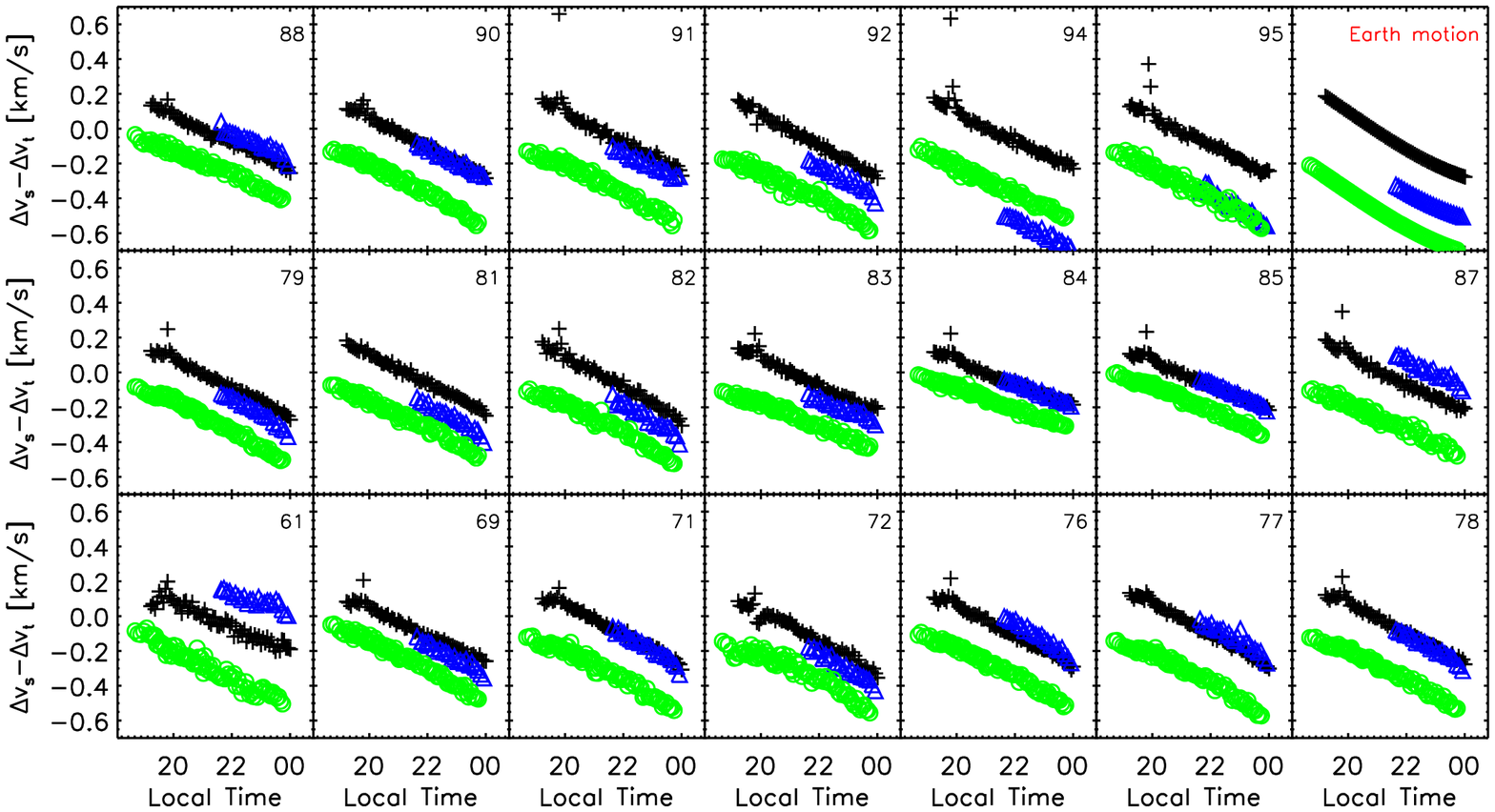}}\\[5mm]
  \resizebox{.97\hsize}{!}{\includegraphics[]{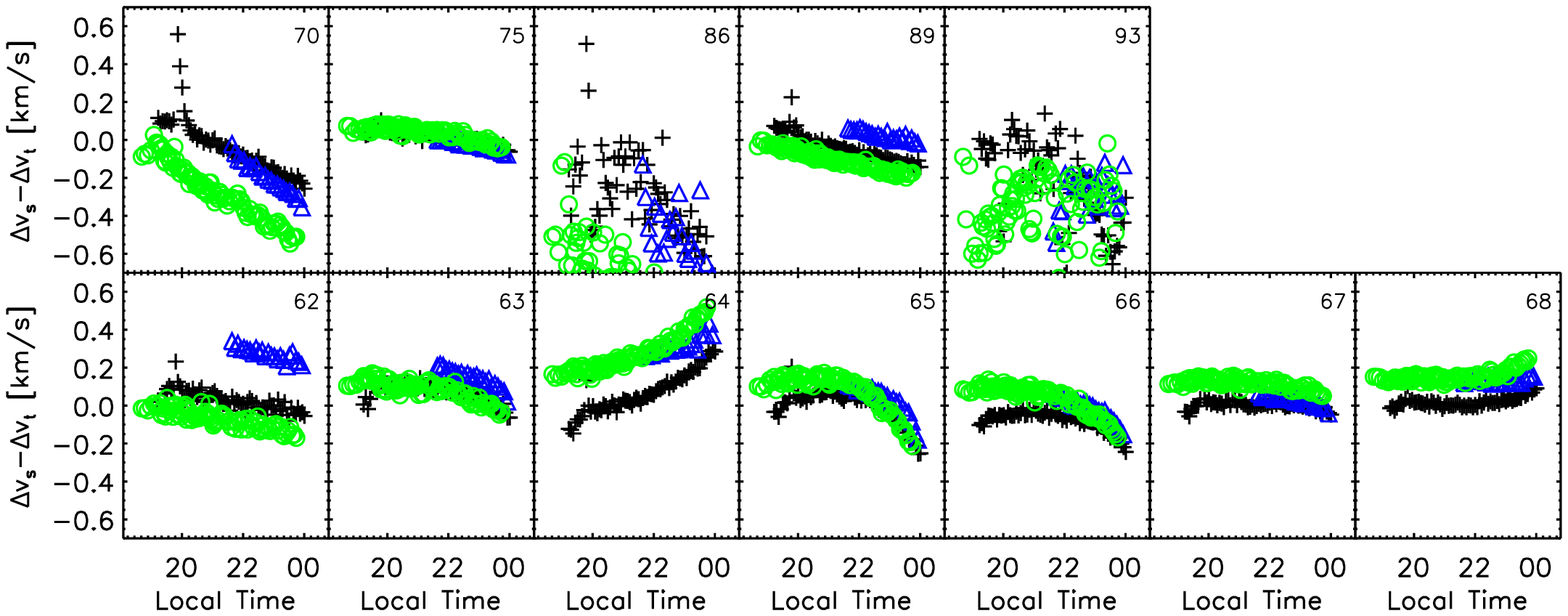}}
  \caption{\label{fig:allorders}Differential radial velocities without
    barycentric corrections for all spectral orders individually.
    Black crosses: first night; blue triangles: second night; green
    circles: third night. Top three panels show orders that resemble
    the pattern of barycentric motion. These orders are compiled in
    Table\,\ref{tab:star_orders}.  The barycentric motion is shown in
    the upper right plot. Lower two panels show spectral orders not
    following barycentric motion.  These orders are affected by
    telluric lines or other contaminants, they are compiled in
    Table\,\ref{tab:tell_orders}.}
\end{figure*}

\begin{table}
  \center
  \caption{\label{tab:star_orders} Spectral orders of UVES that contain mostly stellar absorption features. Orders with relevant emission lines or telluric absorption lines are excluded.}
  \begin{tabular}{lc}
    \hline
    \hline
    \noalign{\smallskip}
    Order \#& Wavelength coverage [nm]\\
    \noalign{\smallskip}
    \hline
    \noalign{\smallskip}
    61 &  992 -- 1008 \\
    69 &  877 -- 891 \\
    71 &  853 -- 865 \\
    72 &  841 -- 853 \\
    76 &  796 -- 809 \\
    77 &  785 -- 798 \\
    78 &  775 -- 788 \\
    79 &  766 -- 778 \\
    81 &  747 -- 759 \\
    82 &  738 -- 750 \\
    83 &  730 -- 739 \\
    84 &  721 -- 730 \\
    85 &  713 -- 721 \\
    87 &  697 -- 705 \\
    88 &  687 -- 699 \\
    90 &  672 -- 683 \\
    91 &  666 -- 674 \\
    92 &  659 -- 666 \\
    94 &  643 -- 654 \\
    95 &  637 -- 647 \\
    \noalign{\smallskip}
    \hline
  \end{tabular}
\end{table}

\begin{table}
  \center
  \caption{\label{tab:tell_orders} Spectral orders of UVES not used for radial velocity calculations. Telluric absorption features or broad stellar features are indicated.}
  \begin{tabular}{lcr}
    \hline
    \hline
    \noalign{\smallskip}
    Order \# & Wavelength coverage [nm] & feature\\
    \noalign{\smallskip}
    \hline
    \noalign{\smallskip}
    62 & 976 -- 991 & H$_2$O \\
    63 & 960 -- 976 & H$_2$O \\
    64 & 945 -- 960 & H$_2$O \\ 
    65 & 931 -- 945 & H$_2$O \\
    66 & 917 -- 931 & H$_2$O \\
    67 & 903 -- 917 & H$_2$O \\
    68 & 890 -- 903 & H$_2$O \\ 
    70 & 865 -- 877 & Ca (stellar)\\
    75 & 808 -- 818 & H$_2$O \\
    80 & 758 -- 767 & O$_2$\\
    86 & 705 -- 713 & He, TiO (stellar)\\
    89 & 681 -- 689 & O$_2$\\
    93 & 652 -- 659 & H$\alpha$ (stellar)\\
   \noalign{\smallskip}
    \hline
  \end{tabular}
\end{table}

In this work, we chose neither to remove nor to mask telluric
features. Instead, we computed the cross correlation function for each
order individually and search for orders that trace the barycentric
velocities. We assume that the velocities calculated from these orders
follow the apparent radial velocity of the target.

\begin{figure}
  \resizebox{.97\hsize}{!}{\includegraphics[]{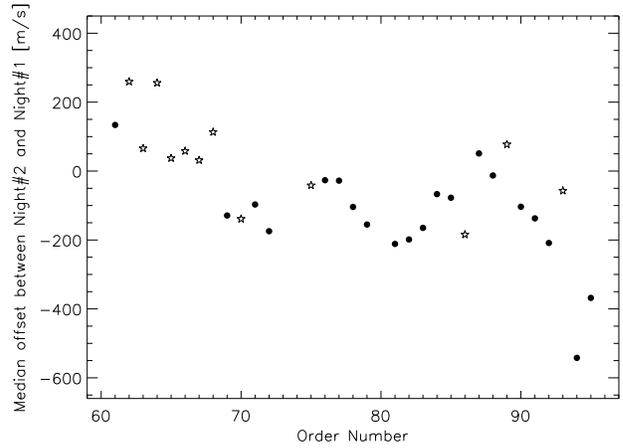}}
  \caption{\label{fig:drift}Offset between the median differential
    radial velocities during night \#2 and night \#1 as a function of
    spectral order (difference between blue triangles and black
    crosses in Fig.\,\ref{fig:allorders}). Filled circles are from
    orders tracing the star's motion, open stars show telluric orders
    (top and bottom panel in Fig.\,\ref{fig:allorders}),
    respectively.}
\end{figure}

With the \emph{relative} radial velocity shift with respect to one
reference frame, we can now calculate the \emph{differential} radial
velocity shift $\Delta v_{\rm rad, star} - \Delta v_{\rm rad,
  telluric} = \Delta v_{\rm s} - \Delta v_{\rm t}$, i.e. the residual
between the relative telluric shift from order \#80 (left panel of
Fig.\,\ref{fig:vrad_raw}) and the relative stellar shift (one order
shown in the right panel of Fig.\,\ref{fig:vrad_raw}). The
differential radial velocities for all orders are shown in
Fig.\,\ref{fig:allorders}. We group the orders into ``stellar'' and
``telluric'' orders based on whether or not they trace the barycentric
motion. The three top panels show ``stellar'' orders following the
apparent velocity introduced by the motion of Earth, which itself is
shown in the upper right panel of Fig.\,\ref{fig:allorders}.  These
orders are compiled with their wavelength coverage in
Table\,\ref{tab:star_orders}. The amplitude of the drift introduced by
the motion of Earth is on the order of 500\,m\,s$^{-1}$ per 4\,h and
clearly dominates the radial velocity signal measured during each
night in the 20 orders shown here. The radial velocity differences
between the three individual nights (sets of black crosses, blue
triangles, and green circles) do not strictly resemble the pattern
expected from barycentric motion.  This is not suprising because the
absolute stability of the spectrograph is not better than a few
hundred m\,s$^{-1}$ from night to night. The differences between the
first and third night are in general closer to the expectations than
the differences between the second and the other two nights. The
differences vary between different orders. We calculate the median of
the differential radial velocity shifts from the first and second
nights and plot their differences in Fig.\,\ref{fig:drift} as a
function of spectral order.  The offset between spectra taken during
the first and the second night varies between $\sim -500$ and
300\,m\,s$^{-1}$ and shows relatively smooth variation with
wavelength, which indicates that the offset is introduced by
inaccuracies of the dispersion relation. Thus, UVES data calibrated
with only one ThAr frame per night can clearly not be used to search
for radial velocity variations on the order of 100\,m\,s$^{-1}$ over
several hours. On the other hand, short-time radial velocity
variations can be traced as will be shown in the following.

The two bottom panels of Fig.\,\ref{fig:allorders} show differential
radial velocities in the spectral orders that do not follow the drift
expected from barycentric motion. These are the orders that are either
affected by large scatter caused by broad features not suitable for
precision velocity measurements (orders \#86 and \#93), or that are
affected by telluric contamination.  Order \#70 was taken out because
it follows the weaker flare at the beginning of night 3 (before
20:00), which is probably caused by the Ca emission line. The spectral
orders shown in the bottom two panels are not used for radial velocity
calculations, they are given in Table\,\ref{tab:tell_orders} together
with the source of contamination.

A comparison between the spectral orders of
Table\,\ref{tab:star_orders} and known telluric absorption features
shows that some spectral regions, e.g., at 690\,nm and at 720\,nm
contain telluric absorption. These orders are likely to be affected by
telluric absorption, too, which can be a problem for radial velocity
work. At least they should be used only with great care. In our case,
Fig.\,\ref{fig:allorders} shows that the stellar signal is dominating
the radial velocity measurement (e.g., in orders 84 and 88), and we
chose to include them in the set of ``stellar'' orders.  Furthermore,
a comparison with the telluric standard spectrum in \citet{Reiners07},
taken with the same instrumental setup, shows that the telluric
features identified in Table\,\ref{tab:tell_orders} represent the
strong features seen from Paranal. No other significant absorption
features (on the order of 20\,\% depth) appears in this atlas. The
relevant parts of the telluric bands at 690\,nm and 720\,nm, for
example, have less than 15\,\% depth, which is much weaker than the
depth of stellar features. We note that classifying more orders that
are suspect to telluric contamination as ``telluric'' has no
significant influence on the results presented in the following.

\begin{figure}
  \resizebox{.97\hsize}{!}{\includegraphics[]{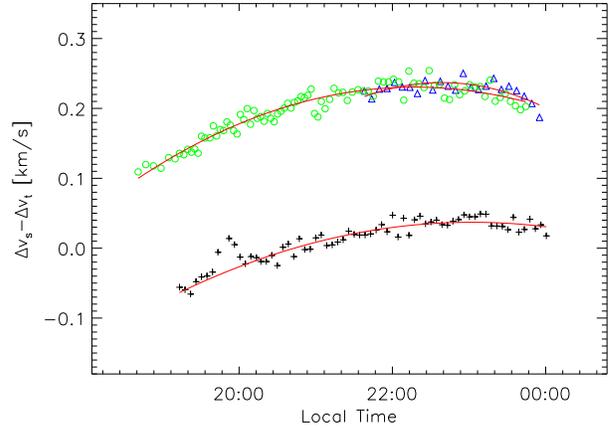}}
  \caption{Differential radial velocities $\Delta v_{\rm s} - \Delta
    v_{\rm t}$ for the three nights versus local time.  Symbols are
    the same as in Fig.\,\ref{fig:vrad_raw}. A second order polynomial
    fit for the three nights is overplotted in red. The giant flare at
    the beginning of the first night was excluded from the fit.}
  \label{fig:vrad_diff}
\end{figure}

Our next step is to apply the barycentric correction and to calculate
the mean differential radial velocity shift from the ``stellar''
orders (top three rows of Fig.\,\ref{fig:allorders} and
Table\,\ref{tab:star_orders}). The mean of the differential velocity
shifts from the 20 orders given in Table\,\ref{tab:star_orders} is
shown in Fig.\,\ref{fig:vrad_diff}.

The differential radial velocity shifts from each of the three nights
all lie on smooth curves. Differential velocity shifts from the second
and the third night show very similar absolute values.  Those from the
first night, however, fall about 200\,m\,s$^{-1}$ below the results of
the other two nights. This discrepancy is probably connected to the
strong drift in the relative radial velocities during the first night.
We checked temperatures during the three nights and found that
relatively large temperature variations occurred during the first
night.  We can speculate that this may be connected to the systematic
offset between the dispersion relations during that night and the
other two nights (Fig.\,\ref{fig:vrad_diff}). However, the temperature
variations cannot explain the large drift in relative velocities seen
in Fig.\,\ref{fig:vrad_raw} \citep[the temperature effect is expected
to be much smaller, see][]{Kaufer07}.

\begin{figure}
  \resizebox{.97\hsize}{!}{\includegraphics[]{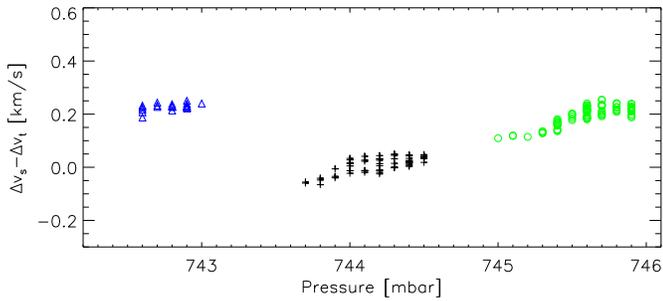}}
  \caption{Differential velocities plotted versus atmospheric pressure
    during the three nights of observation. Symbols are the same as in
    Fig\,\ref{fig:vrad_raw}.}
  \label{fig:vdiff_pressure}
\end{figure}

The smooth variation of differential radial velocity shifts observed
during each of the three nights are all following a very similar
pattern. It is not the main objective of this paper to investigate the
nature of this smooth pattern, because we are mainly interested in
short term variations connected to stellar flares that we can trace in
H$\alpha$.  Nevertheless, we can speculate about the reason of the
smooth changes seen in each of the three nights. The striking
similarity of the nightly variations suggests a reason connected to
the nightly course of observations. One possibility is that some
interference between telluric lines and stellar lines is left in the
orders that we selected above. As explained, this could introduce
changes of radial velocity drift with changing barycentric motion.
However, the barycentric motion is almost monotonous during each night
(upper right box in Fig.\,\ref{fig:allorders}). The nightly velocity
variations consistently show a maximum around 23:00\,h each night,
which makes a connection to telluric lines improbable (although we
cannot completely exclude this option).

A second possibility to explain the intra-night pattern are variations
of atmospheric pressure inside the UVES instrument.  Pressure
variations can alter the dispersion relation and hence affect the
differences between wavelength shifts observed at different spectral
orders. We checked the behaviour of atmospheric pressure during the
three nights; they indeed follow a pattern very similar to the curves
in Fig.\,\ref{fig:vrad_diff}. In Fig.\,\ref{fig:vdiff_pressure}, the
differential velocity shift is plotted as a function of pressure. For
each night, the curves follow a relative well defined line with a
similar dependence on pressure while the absolute pressure values are
very different. The amplitude of pressure variations is consistent
with the expectations from the UVES manual \citep{Kaufer07}.

\subsection{Could the intra-night drift be real?}

A viable explanation for the smooth trend in differential radial
velocities in Fig.\,\ref{fig:vrad_diff} is a change of dispersion
coupled to atmospheric pressure inside the instrument.  However, this
explanation cannot be tested, and the trend may as well be real. If we
assume that the offset between the differential radial velocities,
measured during the first and the other two nights, is instrumental
but the drift comes from real radial velocity variations, the
``detected'' period would be about 2 days or an integer fraction of
that. A potential explanation for such a period in the flare star
CN~Leo are corotating spots producing radial velocity variations with
a semi-amplitude of $K_{\rm s} \sim 100$\,m\,s$^{-1}$. As in higher
mass stars, spots may produce radial velocity variations on the order
of several 10 to 100\,m\,s$^{-1}$ \citep{Bonfils07, Demory07,
  Desort07}.  The rotational period of CN~Leo is not known, but its
surface rotation velocity is about $v\,\sin{i} = 3$\,km\,s$^{-1}$
\citep{RB07}.  \citet{Lacy77} gives a value of $\log{R/{\rm R}_{\sun}}
= -0.84$ ($R = 0.14$\,R$_{\sun}$) for the radius of CN~Leo. Thus, the
(projected) rotational period of CN~Leo is $P/\sin{i} \approx 2$\,d!
This means that rotational modulation as a reason for the radial
velocity changes seen during each of the three nights is not
inconsistent with the expected rotational period of CN~Leo. However,
we have tested the periodicity of H$\alpha$ and other emission lines
and found no sign of a rotational period of 2\,d. This may imply that
the rotational variation is buried under the strong and irregular
flaring activity so that CN~Leo's rotation cannot be detected in
emission lines, but it may also that the rotational period of CN~Leo
is very different from 2\,d.

A third way to generate radial velocity variations as observed in our
data is an object in orbit around CN~Leo. The semi-amplitude caused by
the hypothetical planet would be around 100\,m\,s$^{-1}$, which could
be explained by a planet with a projected mass of $M\sin{i} \sim
0.2$\,M$_{\rm J}$ at a distance of 0.015\,AU. For now, this hypothesis
cannot be falsified with the available data, but seems to be extremely
unlikely given the close match to pressure changes in the instrument
and the coincidence with the estimated rotational period of CN~Leo.

\section{Results}

\begin{figure}
  \resizebox{.97\hsize}{!}{\includegraphics[]{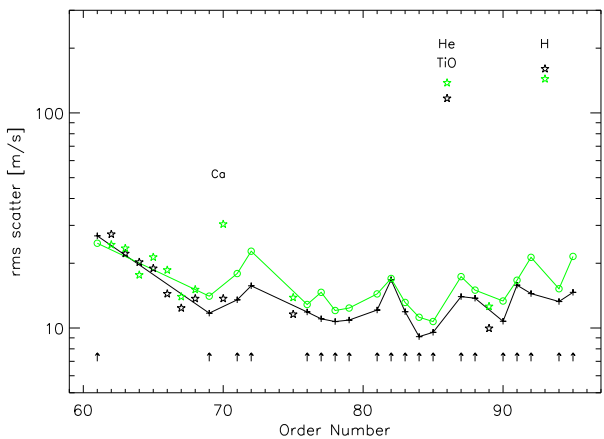}}
  \caption{Scatter of differential radial velocities around the trend
    fitted in Fig.\,\ref{fig:vrad_diff}. Only nights 1 (black crosses)
    and 3 (green circles) are shown. Stars indicate orders that
    predominantly contain telluric features, arrows mark the orders
    used to compute stellar radial velocities (these are the orders
    given in Table\,\ref{tab:star_orders}).}
  \label{fig:scatter}
\end{figure}

The main focus of this work are the short-term variations of the
apparent radial velocity introduced by stellar flares. These flares
have timescales of the order of about half an hour.

We removed the intra-night trend from the differential radial velocity
shifts in Fig.\,\ref{fig:vrad_diff} by applying a second order
polynomial fit (exluding the large flare during the first night). The
resulting final radial velocities are discussed in the following.

\subsection{Radial velocity jitter}

\begin{figure*}
  \resizebox{.97\hsize}{!}{\includegraphics[]{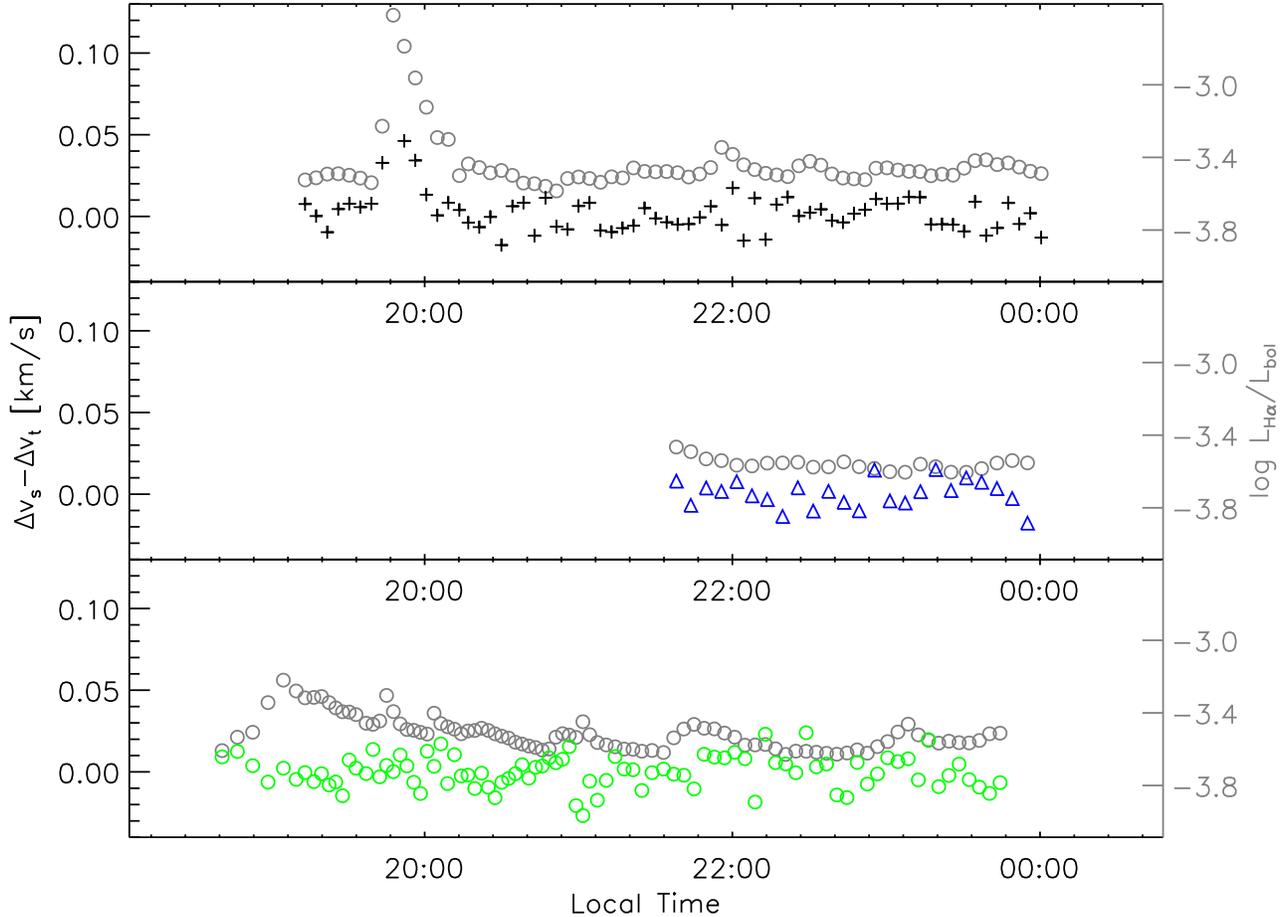}}
  \caption{Radial velocity jitter for the three nights (from top to
    bottom, symbols as above) together with normalized H$\alpha$
    luminosity (grey circles). The polynomials shown in
    Fig.\,\ref{fig:vrad_diff} are subtracted from the differential
    radial velocities.}
  \label{fig:jitter_Halpha}
\end{figure*}

Fig.\,\ref{fig:scatter} displays the rms scatter for all UVES orders
that are covered by our setup. Again, the six spectra taken during the
huge flare were neglected. Only nights 1 and 3 are shown because much
less data is available for the second night. Spectral orders affected
by telluric lines (Table\,\ref{tab:tell_orders}) are marked with a
star. Symbols as in the figures above, i.e. black crosses for night 1
and green circles for night 3, are used for the other orders covering
mainly stellar features (Table\,\ref{tab:star_orders}). These orders
are connected by black and green lines for the first and third night,
respectively.  Interestingly, the rms scatter is decreasing among the
lowest orders (longest wavelengths) from values around 30\,m/s in
order \#61 down to values around 10\,m/s in order \#69. The reddest
order is virtually free of telluric features but contains many strong
stellar lines due to molecular FeH.  The next 7 orders predominantly
contain absorption due to atmospheric ${\rm H_2O}$. The trend of
higher scatter in the reddest orders is probably caused by the quality
of the wavelength solution, which becomes worse towards the near
infrared because the ThAr calibration lamp does not provide as many
lines as in the bluer spectral orders.

Most of the orders have rms values between 10 and 20\,m\,s$^{-1}$.
Several orders show relatively high rms values.  This ``scatter'' is
correlated with H$\alpha$ emission (see next section) and is caused by
chromospheric emission lines contained in these orders. The most
prominent lines that affect radial velocity measurements are the lines
of the Ca infrared triplet at 849.8\,nm, 854.2\,nm, and 866.2\,nm
(orders \#72, \#71, and \#70), two He\,\textsc{I} lines at 706.5\,nm
(order \#86) and 667.8\,nm (orders \#91 and \#92), and H$\alpha$
(order \#93). Order \#86 also contains the strong TiO $\gamma$-band.
This very strong feature is a dense forest of blended lines generating
wide correlation functions, which are part of the reason for the
relatively large scatter in radial velocity.

In Fig.\,\ref{fig:jitter_Halpha}, the mean differential radial
velocity shifts calculated from all orders in
Table\,\ref{tab:star_orders} are shown after substracting the smooth
trend (Fig.\,\ref{fig:vrad_diff}). The first night is plotted in the
top panel, the second in the middle, and the third night in the bottom
panel. In the same figure, the values of normalized H$\alpha$ emission
$\log{L_{\rm H\alpha}/L_{\rm bol}}$ are plotted as grey circles.

\subsubsection{The huge flare}

In Fig.\,\ref{fig:jitter_Halpha}, the strong flare that occurred at the
beginning of the first night is clearly visible as a spike in radial
velocity. Spectra of this huge event are discussed in detail in
\citet{Fuhrmeister08}; such a flare is obvious in the spectra of M
dwarfs. In the spectrum at flare peak, the effect on radial velocity
is as large as several hundred m\,s$^{-1}$ and it shows the opposite
sign ($\Delta v_{\rm s} - \Delta v_{\rm t} = -660$\,m\,s$^{-1}$, not
shown in Fig.\,\ref{fig:jitter_Halpha}).  The exact value of the
radial velocity shift during flare peak has nothing to do with the
star's radial velocity (it may reflect the geometry and dynamics of
the flare). We have not included this point in
Fig.\,\ref{fig:jitter_Halpha}, but we note that its radial velocity is
an order of magnitude larger than in all other spectra.  After the
flare, the radial velocity shift diminishes together with H$\alpha$
luminosity.

In a search for radial velocity variations, the spectra affected by a
flare as huge as this one are easily distinguished from ``quiet''
spectra. In a planet search program, they should not be used to detect
radial velocity variations.

\subsubsection{Jitter outside the huge flare}

The radial velocities outside the huge flare show no obvious trend or
period. The rms scatter for the three nights is 8.2\,m\,s$^{-1}$ for
the first night, 8.4\,m\,s$^{-1}$ for the second night, and
9.7\,m\,s$^{-1}$ for the third night. The overall jitter for all three
nights taken together is 9.0\,m\,s$^{-1}$. The meaning of this result
is twofold: a) For the purpose of this investigation, the accuracy to
which differential radial velocity variations are determined is on the
order of 10\,m\,s$^{-1}$; b) outside the giant flare, the still
relatively strong flaring activity ($\la 0.4$~dex in $\log{L_{\rm
    H\alpha}/L_{\rm bol}}$) on CN~Leo has no effect on radial velocity
measurements in excess of 10\,m\,s\,$^{-1}$ as long as spectral orders
are used that avoid strong chromospheric emission lines.

\subsection{Correlation with activity}

\begin{figure}
  \resizebox{.97\hsize}{!}{\includegraphics[]{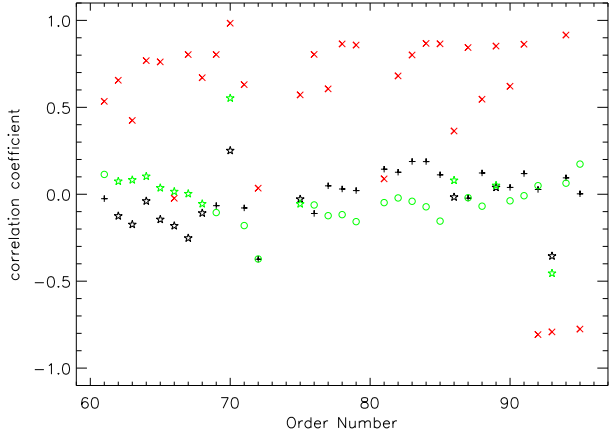}}
  \caption{Correlation coefficients between the radial velocity shifts
    and normalized H$\alpha$ luminosities. Black crosses ($+$) and
    green circles show correlation coefficients for each order
    containing mainly stellar features during the first and third
    nights, respectively (Table\,\ref{tab:star_orders}). Stars show
    orders of Table\,\ref{tab:tell_orders}.  As red crosses
    ($\times$), the correlation coefficients calculated for the first
    night are shown including the spectra taken during the huge
    flare.}
  \label{fig:correlation}
\end{figure}

In the calculation of the rms jitter above, we excluded the spectra
taken during the huge flare. This flare event does show influence on
the radial velocity curve, but flares of this strength pose no danger
to radial velocity surveys because they are obvious in the spectra;
they must not be used for planet searches. However, the data on
H$\alpha$ emission shows that a number of smaller flaring events
occurred during the three nights. Are the smaller flares, which are
not easily detectable in the spectra and will contribute to radial
velocity measurements, affecting the radial velocity measurements?
From a first glance, the curves of radial velocity and H$\alpha$
luminosity in Fig.\,\ref{fig:jitter_Halpha} show no striking
similarity. To check the correlation between radial velocity and
H$\alpha$ activity, we computed the linear Pearson correlation
coefficients (IDL procedure \texttt{correlate}) between normalized
H$\alpha$ luminosity and differential radial velocity.

The results for each spectral order are shown in
Fig.\,\ref{fig:correlation}. Symbols are the same as in
Fig.\,\ref{fig:scatter}. Data for the first night are computed without
the six spectra taken during the strong flare. In most of the spectral
orders, the absolute value of the correlation coefficient is small ($<
0.3$) indicating no or very weak correlation between H$\alpha$
emission and radial velocity.  Thus, flaring shows no obvious effect
on the radial velocities as long as $\Delta v$ is measured in spectral
orders free of chromospheric emission lines. On the other hand,
spectral orders containing prominent emission lines, namely orders
\#70, \#72, and \#93, exhibit rather high (absolute) correlation
coefficients.  Here, activity strongly affects radial velocity
measurements through chromospheric emission in the Ca, He, and
H$\alpha$ lines, although small flares cannot immediately be detected
in the spectra themselves. These orders must to be omitted when
searching for radial velocity variations in active M dwarfs.

In Fig.\,\ref{fig:correlation}, we also show the correlation
coefficients calculated from the entire first night, i.e., including
the huge flare (plotted in red). As mentioned above, this flare
substantially affected the whole spectrum including orders far away
from H$\alpha$ \citep[see][]{Fuhrmeister08}. The red crosses in
Fig.\,\ref{fig:correlation} demonstrate that almost all spectral
orders -- including the ones containing mostly telluric lines -- now
show strong correlation with H$\alpha$ emission. Thus, the radial
velocity shift measured during this flare clearly is a systematic
effect and not due to motions of the flare plasma.  Spectra taken
during flare events similar to this one therefore carry little
information about the real radial velocity of the star (on the order
of a few hundred m\,s$^{-1}$).  Interestingly, the two orders
containing the H$\alpha$ line show an anti-correlation with H$\alpha$
emission. An anti-correlation between radial velocity and H$\alpha$
luminosity was also measured at higher accuracy in a much less active
star by \cite{Kuerster03}.

\section{Summary}

We have used UVES observations of the very active M dwarf CN~Leo
during three nights investigating the effect of flares on short-term
radial velocity determinations. Radial velocities were measured
relative to telluric absorption bands because no bracketing ThAr
exposures or calibrations through an iodine cell were available. With
this method, we could remove strong absolute shifts that occurred
during the nights.  Remaining intra-night drifts may be explained by
pressure variations changing the dispersion of the spectrograph or by
corotating spots on the surface of CN~Leo.  These smooth trends were
neglected in our analysis.

During two nights, data were taken continuously for more than four
hours. This allowed to investigate short-term variations of radial
velocity. Many flaring events were observed during that time. The
timescale of flares is less than an hour and several flares are
covered by our data.

At the beginning of the first night, a huge flare occurred that
dramatically changed the appearance of the whole spectrum. This flare
had strong effect on the measured radial velocities in all spectral
orders. Even spectral orders that predominantly contain telluric lines
show radial velocity shifts on the order of a few hundred m\,s$^{-1}$.
This dramatic kind of flaring events, however, is not a general
problem for radial velocity surveys, because such an event is easily
identified in the spectrum. In this case, the flare limits the radial
velocity stability to $\sim 500$\,m\,s$^{-1}$ in the peak spectrum,
and to $\sim 50$\,m\,s$^{-1}$ directly after the peak. Such events
must be removed from the analysis.

Outside the huge flare, spectral orders containing prominent emission
lines show large scatter. Radial velocities measured in the order
containing the H$\alpha$ line are anti-correlated to H$\alpha$
emission strength. This was also reported in the very inactive M dwarf
Gl~699 by \citet{Kuerster03}.

In the orders free of prominent emission lines, the radial velocity
jitter is below 10\,m\,s$^{-1}$. Outside the huge flare, CN~Leo still
showed flaring activity that enhanced the normalized H$\alpha$ flux by
$\la 0.4$\,dex. Despite this rather strong activity, no correlation
was found between activity and radial velocity.

Even in one of the most active known M dwarfs, flares produced during
average stellar activity have no effect on high precision radial
velocities on the level of 10\,m\,s$^{-1}$ if one chooses spectral
orders free from emission lines. This means that the influence of
flares -- the most obvious effect of stellar activity on the spectra
of M dwarfs -- is negligible in planet searches at least down to
several m\,s$^{-1}$.  Even in the presence of flaring, it may be
possible to detect an earth-like planet inside the habitable zone of a
late M dwarf.

\begin{acknowledgements}
  I thank Andreas Seifahrt and Jacob Bean for many enlightening
  discussions on the subject, and Uwe Wolter and Carolin Liefke for
  providing the reduced data set. Research funding is acknowledged
  from the DFG under an Emmy Noether Fellowship (RE 1664/4-1). An
  encouraging and very helpful report by an anonymous referee is
  thankfully acknowledged.
\end{acknowledgements}


\begin{thebibliography}{}

\bibitem[Balthasar et al., 1982]{Balthasar82}Balthasar, H., Thiele,
  U., \& W\"ohl, H., 1982, A\&A, 114, 357
\bibitem[von Bloh et al., 2007]{vonBloh07}von Bloh, W., Bounama, C.,
  Cuntz, M., Franck, S., 2007, A\&A, 476, 1365
\bibitem[Bonfils et al., 2007]{Bonfils07}Bonfils, X., Mayor, M.,
  Delfosse, X., et al., 2007, A\&A, 474, 293
\bibitem[Butler et al., 1996]{Butler96}Butler, R.P., Marcy, G.W.,
  Williams, E., McCarthy, C., Dosanjh, P., \& Vogt, S.S., 1996, PASP,
  108, 500
\bibitem[Demory et al., 2007]{Demory07}Demory, B.-O., Gillon, M.,
  Barman, T., et al., 2007, A\&A, 475, 1125
\bibitem[Desort et al., 2007]{Desort07}Desort, M., Lagrange, A.-M.,
  Galland, F., Udry, S., \& Mayor, M., 2007, A\&A, 473, 983
\bibitem[Fuhrmeister et al., 2008]{Fuhrmeister08}Fuhrmeister, B.,
  Liefke, C., Schmitt J.H.M.M., \& Reiners, A., A\&A, in press
\bibitem[Kaufer et al., 2007]{Kaufer07}Kaufer, A., D'Odorico, S.,
  Kaper, L., Ledoux, C., 2007, UVES User Manual, ESO
\bibitem[K\"urster et al., 2003]{Kuerster03}K\"urster, M., Endl, M.,
  Rouesnel, F., et al., 2003, A\&A, 403, 1077
\bibitem[Lacy, 1977]{Lacy77}Lacy, C.H., 1977, ApJSS, 34, 479
\bibitem[Pepe et al., 2002]{Pepe02}Pepe, F., Mayor, M., Galland, F.,
  Naef, D., Queloz, D., Santos, N.C., Udry, S., \& Burnet, M., 2002,
  A\&A, 388, 632
\bibitem[Reiners \& Basri, 2007]{RB07}Reiners, A., \& Basri, G., 2007,
  \apj, 656, 1121
\bibitem[Reiners et al., 2007]{Reiners07}Reiners, A., Homeier, d.,
  Hauschildt, P.H., \& Allard, F., 2007, A\&A, 473, 245
\bibitem[Saar \& Donahue, 1997]{Saar97}Saar., S.H., \& Donahue, R.A.,
  1997, \apj, 485, 319
\bibitem[Selsis et al., 2007]{Selsis07}Selsis, F., Kasting, J.F.,
  Levrard, B., Paillet, J., Ribas, I., \& Delfosse, X., 2007, A\&A,
  476, 1373
\bibitem[Short \& Doyle, 1998]{Short98}Short, C.I., \& Doyle, J.G.,
  1998, A\&A, 336, 613
\bibitem[Tarter et al., 2007]{Tarter07}Tarter, J.C., Backus, P.R.,
  Mancinelli, R.L., et al., 2007, Astrobiology, 7, 30
\bibitem[West et al., 2004]{West04}West, A.A., Hawley, S.L.,
Walkowicz, L.M., Covey, K.R., Silvestri, N.M., and 6 authors
2004, \aj, 128, 426

\end{thebibliography}
\end{document}